# Continuous-wave operation and 10-Gb/s direct modulation of InAsP/InP sub-wavelength nanowire laser on silicon photonic crystal

Masato Takiguchi,[1,2,a)] Atsushi Yokoo,[1,2] Kengo Nozaki,[1,2] Muhammad Danang Birowosuto,[1,2, b)] Kouta Tateno,[1,2] Guoqiang Zhang,[1,2] Eiichi Kuramochi,[1,2] Akihiko Shinya,[1,2] and Masaya Notomi,[1,2,c)]

[1] *NTT Basic Research Laboratories, NTT Corp. 3-1, Morinosato Wakamiya Atsugi, Kanagawa 243-0198, Japan.*

[2] *NTT Nanophotonics Center, NTT Corp. 3-1, Morinosato Wakamiya Atsugi, Kanagawa 243-0198, Japan.*

We demonstrated sub-wavelength (~111 nm diameter) single nanowire (NW) continuous wave (CW) lasers on silicon photonic crystal in the telecom-band with direct modulation at 10 Gb/s by optical pumping at cryogenic temperatures. To estimate the small signal response and pseudo-random bit sequence (PRBS) modulation of our CW lasers, we employed a new signal detection technique that employs a superconducting single photon detector and a time-correlated single photon counting module. The results showed that our NW laser was unambiguously modulated at above 10 Gb/s and an open eye pattern was obtained. This is the first demonstration of a telecom-band CW NW laser with high-speed PRBS modulation.

a) Electronic mail: takiguchi.masato@lab.ntt.co.jp

b) *Present address: CINTRA UMI CNRS/NTU/THALES 3288, Research Techno Plaza, 50 Nanyang Drive, Border X Block, Level 6, Singapore 637553, Singapore*

c) Electronic mail: notomi.masaya@lab.ntt.co.jp

## INTRODUCTION

Nano-lasers are promising devices with which to realize ultracompact coherent light sources with low power consumption for intra-chip interconnects and photonic networks-on-chip [1], sensing [2] and cavity quantum electrodynamics [3]. Many kinds of nano-lasers have already been investigated [4-6]. A nanowire (NW) laser is one of the strongest candidates because NWs have a unique sub-wavelength functional structure and are composed of attractive functional materials. NWs are not damaged during the fabrication process unlike samples fabricated by electron beam lithography, dry etching and focused ion beam processes, because the NWs are spontaneously grown from a substrate by a bottom-up approach without the need for lithography. Moreover, the NW emission wavelength can be tuned easily by changing the NW material and composition. In addition, NWs can provide high functionality because quantum disks [7], quantum dots [8, 9] and p-i-n structures [10, 11] can be embedded in their structure. Therefore, NW lasers are very attractive as on-chip light sources.

Pulsed and continuous wave (CW) NW lasers have already been demonstrated at visible and near visible wavelengths [12-19], but the NWs used for lasing were relatively thick and long, and so incapable of fully enjoying the merits of NWs. To realize tiny and efficient single NW emitters, the NWs should be implemented with optical nanocavities to obtain strong light confinement. Recently, pulse and CW plasmon lasers based on sub-wavelength single NWs (diameter : ~100 nm) have also been demonstrated using a metal surface [20-22]. However, there is still a technical problem regarding integration and optical connection for these plasmonic structures. In addition, plasmonic cavities have a huge intrinsic loss that might limit the performance of light emitting devices, especially lasers. On the other hand, photonic crystals have a suitable structure for lasers in optical circuits because it is easy to couple to other devices via waveguides. An NW laser based $TiO_2$ PhC has been demonstrated in the visible region [23].

In the telecom-band region, pulsed GaSb micro-wire lasers have been demonstrated (diameter : ~ 1 μm) at low temperature [24]. We have also reported the pulsed operation of a single telecom-band NW laser by combining a sub-wavelength InAsP/InP NW (diameter : ~ 100 nm) and a Si PhC at 4K [25, 26]. This device was integrated in a Si PhC, and a high Q cavity can be easily achieved. This laser can be integrated with other nano-devices such as detectors [27] with a small optical loss by using a PhC system. Therefore, our single NW laser based on a Si platform is unique and promising for future on-chip applications. However telecom-band CW NW lasers have not yet been demonstrated.

High-speed nanolaser modulation is important for signal processing. Dynamic high-speed modulation has been reported using nano and micro-size lasers, such as VCSELs, micro disk lasers, PhC LEDs and PhC lasers [28-36], but there have been very few reports on NW lasers. To demonstrate fast direct modulation for bit stream data, the devices should achieve CW lasing operation, which is generally hard for NW lasers. In fact, the ultrafast modulation of NW lasers [37, 38] was reported to observe the carrier dynamics at wavelengths of about 520 nm by double femtosecond pulse excitation, but these lasers operated only under a pulsed condition, and thus they did not show continuous modulation for repetitive pulses. High-speed repetitive operation is the most important feature for any signal modulation application and is a prerequisite for CW operation. To demonstrate this, it is essential to demonstrate small-signal modulation for sinusoidal signals and large-signal modulation for bit stream patterns, such as pseudo-random bit sequence (PRBS) modulation. As far as we know, there has been no report of such modulation for NW lasers.

The observation of high-speed repetitive modulation for NW lasers is also hindered by its small output intensity, which is generally too small to be measured with a conventional setup combining a digital oscilloscope and a low-noise erbium-doped fiber amplifier (EDFA) [36] or with a high-speed streak camera [30]. In this study, we employed a powerful and simple measurement technique using superconducting single photon detectors (SSPDs) and a time-correlated single photon counting (TCSPC) module. To demonstrate high-speed modulation for repetitive signals, we estimated an eye diagram, which is the commonly used measure of laser modulation. The eye diagram was obtained from a PRBS signal measured with SSPDs and a TCSPC module. This is a unique analysis approach, and this paper describes its first demonstration.

In this study, first, we show the light-in vs light-out (L-L) property and photon statistics of our sample to estimate if it can work as a CW single NW laser. Here we show a kink structure in the L-L plot and a transition from thermal light to coherent light. If the *β* values of the lasers are high, their lasing threshold becomes more ambiguous. Therefore, a photon correlation measurement could prove whether or not the samples truly achieve lasing. Second, we modulated the emission light by controlling the pump light with a sinusoidal signal and a non-return-to-zero (NRZ) PRBS signal. Our method for measuring a modulated signal is unique in that it uses SSPDs and a TCSPC module. This method overcomes the problem of weak signal detection. By analyzing this result, we also obtain a 10 Gb/s modulation average eye pattern. This is the first demonstration of the modulation of a single NW laser and also strong evidence for CW laser oscillation in the telecommunication band.

## SAMPLE

InP-based NW were grown by employing a vapor-liquid-solid (VLS) mechanism in a low-pressure (76 Torr) horizontal metalorganic vapor phase epitaxy (MOVPE) reactor [39,40]. The substrates were InP(111)B. The catalysts were Au particles (40 nm in diameter) obtained from Au colloids, which were dispersed on the substrate before growth. The growth temperature was 430℃. To embed multiple quantum wells (MQWs) in the NW, we repeated the sequence of InP growth (5 s), InAsP growth (0.2 s) and growth interruption (5 s) 200 times. Energy-dispersive X-ray spectroscopy showed that the As content ranged from 0.6 to 0.7 in this repeated region except in the NW edge region, which reveals that the InP layer also contains As (See supplemental material S1). We transferred the NWs to a Si substrate by realizing direct physical contact between the NWs on the InP substrate and the Si substrate. Then, because the NWs were randomly dispersed on the Si substrate, we carefully manipulated one of the NWs and placed it in a trench in a line defect in a two-dimensional hexagonal PhC slab using an atomic force microscope (AFM) manipulation method [41] (FIG. 1). Although the cavity formation mechanism was the same as in our previous experiment [41], we used different NWs to achieve lasing oscillation. The number of MQWs in each NW was increased from 10 to 100 to increase the gain volume, and the polarization of the intrinsic emission of an NW was optimized to the cavity polarization to improve the extraction efficiency by improving the NW's crystal quality. The NW was 2.5 μm long and about 111 nm in diameter (the maximum and minimum diameters were 144 nm and 78 nm, respectively). The Si PhC slab was 200 nm thick, the lattice constant was 370 nm and the air hole diameter was 200 nm. The trench was 115 nm deep and 150 nm wide. Once an NW was located in the trench in the PhC, the refractive index change induced by the NW created a modulated mode-gap cavity [42].

## DEMONSTRATION OF CW LASING

FIG. 2(a) shows the L-L property and FWHM and FIG. 2(b) shows the spectrum of a NW laser under 0.25 and 2.5 mW CW pumping conditions. In this paper, the pump power is defined as the total incident power on the sample surface, therefore, the transmission loss of the measurement equipment is subtracted. Our measurement system is a typical micro-photoluminescence (PL) system. We pumped our sample from the top with a 785 nm CW semiconductor laser and measured our sample in a He-flow cryostat system at 4K. When the pump power was increased, a kink structure appeared in the L-L curve and FWHM. As many reports suggest, this is typical nanolaser behavior [5, 6, 31-33]. These results strongly indicate that CW lasing oscillation was achieved. Here we estimated that the threshold was 1.4 mW by linear fitting (blue line in FIG. 2(a)). The estimated Q factor of the PhC cavity was 9200 (linewidth of 0.145 nm) at the kink structure in the L-L plot. Theoretically, the Q value of these types of cavities can exceed 100,000 [42]. In reality, the fabricated cavities could not achieve such a high Q due to the absorption loss, fabrication error, roughness and inhomogeneous structure of the NWs. However, the Q value for such cavities can exceed several thousand. The Q value was almost the same as that of other modulated mode-gap cavities for nanolasers [33]. When the pump power was above the lasing threshold, we could directly observe a strong lasing emission with an infrared CCD camera (FIG. 2(c)). We could measure the strong scattered light around the PhC structure and some fringes around the NW. In this case, we obtained an output power of 20-30 nW at the end of a fiber in our micro PL system. Although the experimental threshold power is 1.4 mW, only a small part of this value was absorbed because there was a large mismatch in the focused beam and the NW and the NW absorbance is small. Hence, here we present a rough estimation of the effective threshold in terms of the absorbed power. The illuminated NW area is 3.9% of the focused pump beam area (assuming a Gaussian spot with a diameter of 2 μm), calculated by $\int_{NW} I \, dS / \int_{ALL} I \, dS$. Next, we estimated that ~9.5% is absorbed by the nanowire (assuming an absorption coefficient of about 33000 cm-1 at 785 nm). Consequently, we can deduce that the absorbed power is 0.37% of the total incident power. As a result, the effective lasing threshold is 5.3 μW in terms of the absorbed power. This threshold is close to that of another PhC laser at cryogenic temperatures [43].

## PHOTON CORRELATION MEASUREMENT

To further confirm that lasing oscillation was achieved we performed a photon correlation measurement using a Hanbury Brown and Twiss setup [44,45]. Generally, when the pump power is below the lasing threshold, a bunching signal should be observed ($g^{(2)}(0) > 1$), because the intensity fluctuation (thermal light) is large below the lasing threshold. On the other hand, when the pump power is sufficiently above the lasing threshold, $g^{(2)}(0)$ becomes unity, because the coherent light becomes dominant. It has been well established that this transition of photon statics is considered as a clear signature of lasing for nanolasers. In our setup (FIG. 3(a)), we used a fiber beam splitter, two SSPDs (quantum efficiency of over 10% and jitter time of 25 ps), and a TCSPC module. We also installed a longpass filter and a tunable narrow bandpass filter in front of the detectors to eliminate the reflection of the CW pump laser and the background emission (ASE noise from the wide spectrum of an NW) and an ND filter to attenuate the emission until a photon counting rate of around $10^5$ was realized. First an SSPD generates a start signal and a second SSPD generates a stop signal. A TCSPC module collects the photon's arrival time between the start signal and the stop signal. Thus we can obtain photon correlation spectra. We summarize the bunching spectra for different pump power conditions in FIG. 3(b). This shows a clear lasing transition from a non-lasing condition. At near the lasing threshold, we obtained $g^{(2)}(0) = 1.5$. The reduction of $g^{(2)}(0)$ from 1.8 to 1.5 means that stimulated emission had increased and lasing oscillation had begun. Finally, $g^{(2)}(0)$ becomes unity above the lasing threshold because the emission light becomes coherent light. Combining this correlation measurement and the aforementioned CW measurement, we are now confident that CW lasing is achieved in our NW lasers. This is the first demonstration of CW lasing for sub-wavelength NWs.

In addition to the demonstration of lasing, it has been pointed out that this photon correlation measurement can also provide information about the relaxation oscillation frequency under CW pumping conditions [46-48]. With conventional lasers, relaxation oscillation can be observed when the pump signal is modulated because there is a time delay between the injected carrier and the output photon (this delay is called the turn-on delay time) and it affects the photon number [49]. In general, relaxation oscillation limits the modulation speed of semiconductor lasers. On the other hand, even if the signal level is very weak, in a photon correlation measurement for nanolasers, relaxation oscillation can be observed as a periodic modulation in the second-order correlation function $g^{(2)}(t)$ under CW pumping conditions. This has been theoretically predicted [46] and also measured [47,48]. FIG. 3(c) shows an enlarged bunching spectrum that appeared when the pump power was 1.3 mW. In this graph, we can observe small dips on both sides of the bunching spectra. These dips represent modulation due to the relaxation oscillation. This is typical behavior for CW lasing at near the lasing threshold. At this point, the emission light was perturbed by switching between the spontaneous emission state (non-lasing) and the stimulated emission state (lasing). From this experiment, we can find the relaxation oscillation frequency near the threshold (not the maximum relaxation oscillation frequency). If the linewidth of is sufficiently wide, we can measure several modulation peaks [47]. However, in our case we could only measure the first modulated peak, because the peak width is comparable to the $g^{(2)}(t)$ linewidth. To estimate the relaxation frequency, we fitted the spectrum using the fitting function $g^{(2)}(t) = \left(g^{(2)}(0) - 1\right)e^{-\gamma_d|t|}\cos(\omega_r|t|) + 1$, where $\gamma_d$ is the damping rate of the relaxation oscillation, and $\omega_r$ is the oscillation frequency. We estimated $\omega_r/2\pi$ as 2.3 GHz and $\gamma$ as 11.5 GHz by fitting when the pump power was 1.3 mW (FIG. 3(c)). $\omega_r$ and $\gamma_d$ can be written as $\sqrt{\beta\gamma\kappa n_o - (\frac{\kappa}{n_o+1} - \gamma(1 + \beta n_0))/4}$ and $(\frac{\kappa}{n_o+1} - \gamma(1 + \beta n_0))/2$, where κ, γ and $n_0$ are the cavity decay rate, radiative recombination rate, and photon number, respectively [47,48]. Near the threshold, $\omega_r/2\pi$ and $\gamma_d$ became 4.7 and 16 GHz, respectively (κ∼25 GHz, γ∼10 GHz, β∼0.1, See supplemental material S2). Therefore, our experimental value corresponds well with the theoretical value. When the pump power greatly exceeds the lasing threshold, $\omega_r$ becomes faster. The above equation showed that a maximum theoretical value of 22 GHz could be achieved for $\omega_r/2\pi$ when the pump power was higher than the lasing threshold. However, we cannot observe a faster relaxation frequency in a photon correlation measurement because $g^{(2)}(0)$ decreases as the pump power increases.

## DYNAMIC PROPERTY

Since we have achieved CW lasing, we are now able to pursue repetitive signal modulation for our NW lasers. To accomplish this, we measured a directly-modulated signal from our single NW laser to investigate its dynamic property. Here we determined the 3dB frequency bandwidth from small-signal modulation measurements (sinusoidal modulation) and also performed large-signal modulation measurements by using PRBS input data by which we obtained an eye pattern. With a conventional semiconductor laser the frequency response is usually evaluated from small and large signal modulations using a sampling oscilloscope and a network analyzer. However, the emission intensity of our device is too small for us to measure these modulations using such a measurement system, therefore we used SSPDs and a TCSPC module. First, we averaged the detected signal on a computer to obtain a modulated sinusoidal and PRBS signal. And then, to obtain the eye pattern, we superimposed the data over a specific time period (several nanoseconds). Although we cannot estimate the absolute value of the bit error rate with this measurement method, we can estimate it if our device can be modulated from an average eye pattern.

FIG. 4(a) shows a schematic of the setup we used to measure a small signal modulation using SSPDs and a TCSPC module. The CW pump laser (800 nm) was modulated with a lithium niobate (LN) modulator. The LN modulator was operated with a sinusoidal electric signal from a microwave signal generator. Our single NW laser was pumped by the modulated signal, and the signal from the NW laser was measured with an SSPD and a TCSPC module. Here we averaged the modulated signal until the detected signal became clear. To modulate the pump laser above the lasing threshold, 50% of DC bias was added as shown in the inset of FIG. 4(a). FIG. 4 (b) shows measured signals from the NW laser and from the pumping laser at 5, 10, and 12 GHz. The green regions in the figure show the estimated DC level. As shown in FIG. 4 (b), the DC bias exceeded the lasing threshold (the average pump power was 2.5 mW, the peak intensity was 3.4 mW and the DC bias level was 1.7 mW). When the modulation frequency was 5 GHz, we could detect clear modulation signals. We also measured the modulation signal when we increased the frequency to 10 and 12 GHz. However, the modulation signal was not so clear in these cases. When we increased the modulation frequency to 20 GHz, we could not obtain a modulated signal although our SSPD could respond at a faster frequency (FIG. 4 (c)). We summarize these results as the degree of modulation in FIG. 5 (a). The red dots are results obtained with an NW laser, the black dots were obtained with an SSPD, and the blue dots show the normalized ratio, which is defined as $\frac{P_{max}-P_{min}}{P_{max}+P_{min}}$, where $P_{max}$ and $P_{min}$, respectively, are the maximum and minimum intensities of the modulated signal from the bias level. From this result, we found that our SSPD could provide a response of at least 15 GHz. To estimate the NW laser's response more precisely, we normalized the data using the SSPD frequency response. From the normalized ratio, our NW laser can be modulated at about 10 GHz. This speed is fast enough for on-chip data communication [1], although this value is still slower than the value of 22 GHz that we estimated from the relaxation oscillation result in the photon correlation measurement. We believe that the observed response speed might be limited by the thermal effect or gain saturation, which we did not consider in the theoretical model.

Finally we measured an NRZ PRBS modulated signal from our single NW laser to obtain an eye pattern (bit number : $2^7$-1, 10Gb/s). Similar to the small signal modulation measurement, we obtained our eye pattern from an averaged signal using an SSPD and a TCSPC module for several tens of seconds. We modulated the pump laser using a pulse pattern generator (PPG) in the same setup as the above experiment (FIG. 4(a)). We measured the L-L curve by using a modulated signal and compared it with the L-L curve obtained with CW pumping (See supplementary material S3). This result showed that the effective thresholds and L-L curve were the same as those obtained with the CW pumping. This means that PRBS modulation does not affect our lasing property. FIG. 6 shows the time-integrated signals from our NW laser that we obtained when the pump powers were large and small (the average powers were 2.5 and 0.75 mW, respectively). When the pump power was 2.5 mW (above the threshold), we were able to measure a clear 10 Gb/s modulation signal, but when the pump power was 0.75 mW (below the threshold), the device could not be modulated.

We also analyzed the eye pattern of our NW laser (FIG. 7(a)), to estimate whether or not our NW laser works as signal transmission device. FIG. 7(b) shows the eye pattern of our SSPD (the pump light was directly coupled to the SSPD). To create this pattern, we superimposed our measured PRBS data (long-term data for the modulation frequency, ~ 5 ns). FIG. 7 shows that the eye is clearly open and our single NW laser can be directly modulated at 10 GHz. This also shows that we could successfully measure the eye pattern using the SSPD system even when the emission was weak. Although the top part of the eye pattern is slightly noisier than in FIG. 7(b), this might be because of the intensity fluctuation caused by the heating effect under lasing conditions and the effect of relaxation oscillation. This kind of intensity fluctuation is sometimes observed when the pump power is strong in our setup. Our background noise level did not increase in our open eye pattern, even after we averaged the signal. This implies that our signal bit error is small. Consequently, this measurement shows that our NW laser can transmit a 10 Gb/s NRZ signal with high speed modulation.

## SUMMARY

In this study, we described directly-modulated sub-wavelength nanowire lasers on silicon. We demonstrated the CW lasing of a telecom-band NW for the first time. To estimate the dynamic property of our single NW laser, we pumped it using a sinusoidal and PRBS modulated pump laser. By analyzing the NRZ PRBS signals from our NW laser, we observed an open 10 Gb/s eye pattern. This means our NW laser was unambiguously modulated at 10 Gb/s. We also estimated a relaxation frequency of a few GHz from a photon correlation measurement near the lasing threshold, and it should be faster above the lasing threshold. This was consistent with our modulation measurement.

In the present experiment, we demonstrated a CW telecom band NW laser at low temperature. We must solve some problems if we are to realize a room temperature NW laser. First, the NW may not come into full contact with the Si surface, and so the thermal conductance could be low. Therefore, the use of an embedded structure [50, 51] may be an effective way to improve thermal conductance, Q factor, and nonradiative recombination at the NW surface. Second, we should also consider adopting a NW with a larger gain volume.

## ACKNOWLEDGEMENTS

This work was supported by JSPS KAKENHI Grant Number 15H05735.

## REFERNCES

**FIGURES**

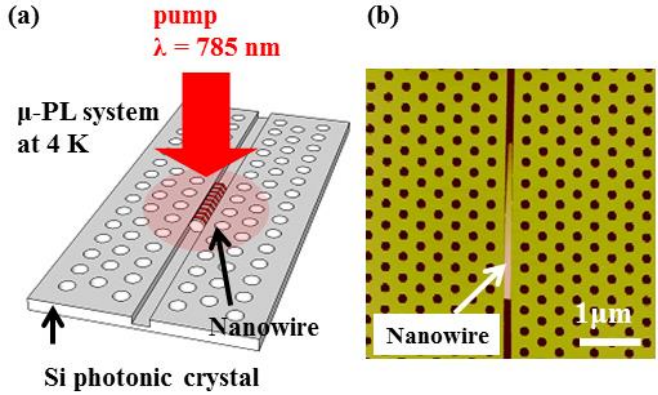

FIG. 1. (a) Schematic of a single NW laser on Si photonic crystal. Our sample is in a cryostat (4K). A NW is located in an air trench in the PhC. The sample is excited from the top. The average diameter is 111 nm (min of 78 nm; max of 144 nm) and the length is 2.5 μm. The NW was set in an air trench (114-nm deep; 150-nm wide) in a Si PhC using an AFM. (b) AFM image of an NW in a PhC trench.

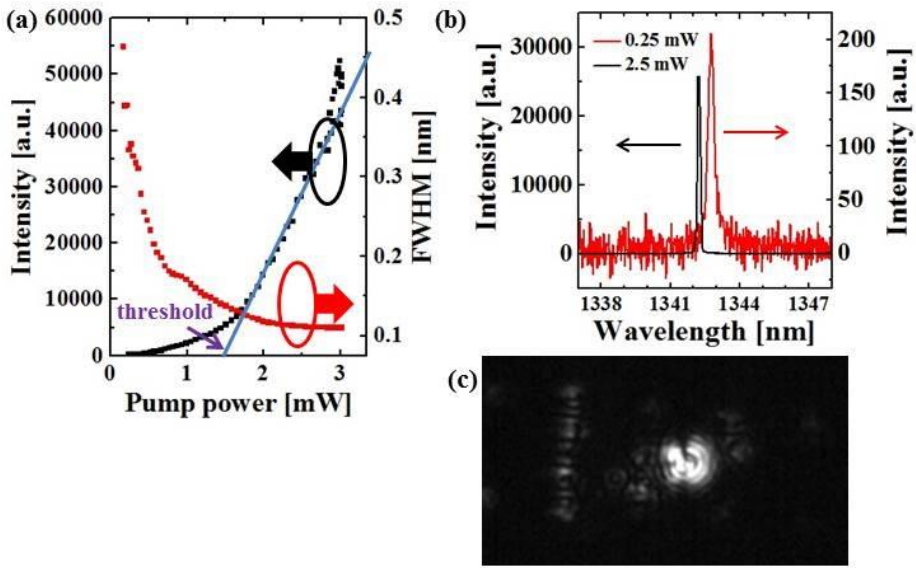

FIG. 2. (a) L-L curve and full width at half maximum of pump power. The black dots are intensity and the red dots are the cavity line width. The blue line is a fitting line to determine the lasing threshold. (b) Spectrum of an NW on a PhC at 0.25 and 2.5 mW. (c) IR image from the top under strong pumping conditions without a back light (illumination).

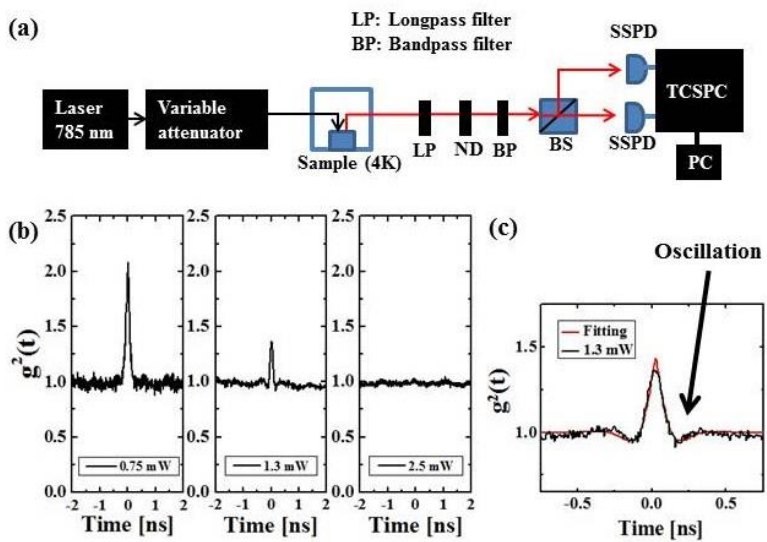

FIG. 3. (a) Schematic of our measurement setup for photon correlation measurement. (b) Photon correlation measurement for different pump powers (0.75, 1.3 and 2.5 mW). (c) Fitting for photon correlation measurement at 1.3 mW.

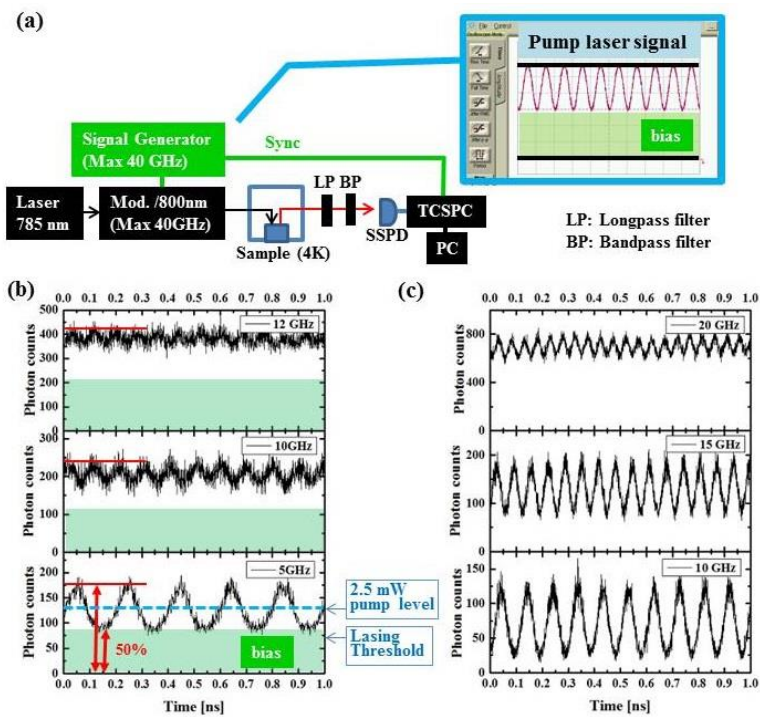

FIG. 4. (a) Schematic of our measurement setup for modulation measurement. Inset picture shows the modulated signal of the pump laser measured with a fast photo-diode. (b) Sinusoidal modulated signals of a NW laser (5, 10 and 12 GHz). The green region is the DC bias. (c) Sinusoidal modulated signals of the pump laser without the bias level. This represents the response of the SSPD (10, 15 and 20 GHz).

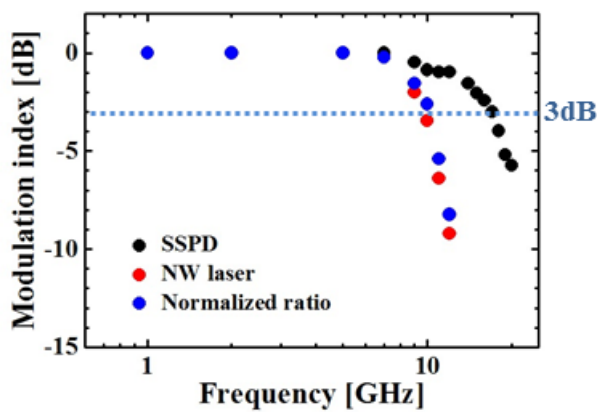

FIG. 5. Comparison of modulation index of our NW laser (red dots), an SSPD (black dots), and the normalized ratio (blue dots).

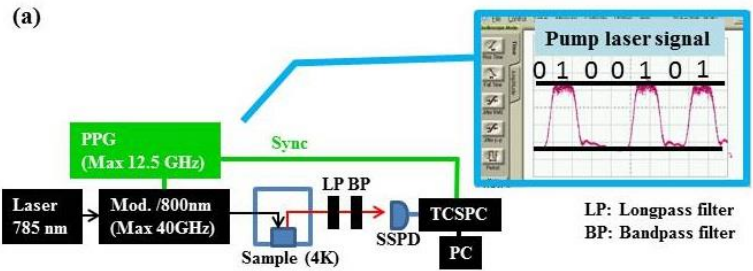

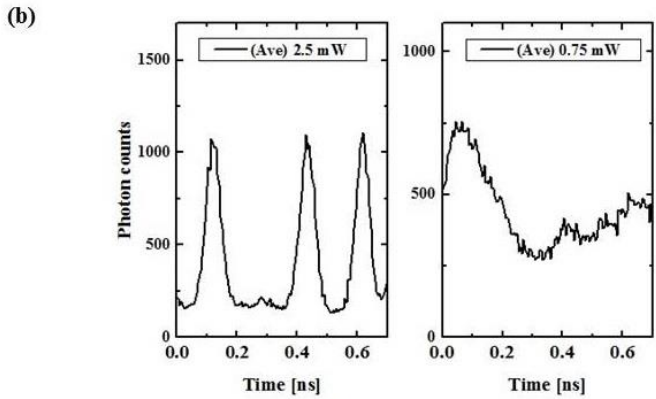

FIG. 6. (a) Schematic of our measurement setup for modulation measurement. Inset picture shows the modulated signal of the pump laser measured with a fast photo-diode. (b) 10-Gb/s pseudo-random bit sequences from the NW laser (right: 2.5 mW, left: 0.75 mW).

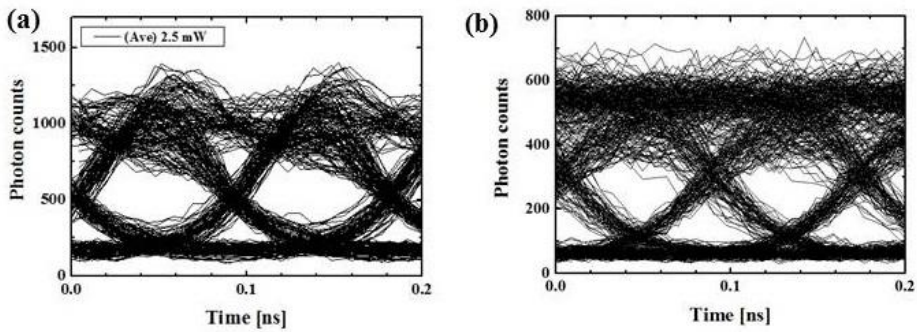

FIG. 7. (a) Eye diagram of a NW laser (10 Gb/s) (b) Eye diagram of an SSPD (10 Gb/s)

# Supplementary material for Continuous-wave operation and 10-Gb/s direct modulation of InAsP/InP sub-wavelength nanowire laser on silicon photonic crystal

Masato Takiguchi,[1,2,a)], Atsushi Yokoo,[1,2], Kengo Nozaki,[1,2], Muhammad Danang Birowosuto,[1,2,b)], Guoqiang Zhang,[1,2], Kouta Tateno,[1,2], Eiichi Kuramochi,[1,2], Akihiko Shinya,[1,2], and Masaya Notomi,[1,2,c)]

[1] *NTT Basic Research Laboratories, NTT Corp. 3-1, Morinosato Wakamiya Atsugi, Kanagawa 243-0198, Japan.*

[2] *NTT Nanophotonics Center, NTT Corp. 3-1, Morinosato Wakamiya Atsugi, Kanagawa 243-0198, Japan.*

## S1. Energy-dispersive X-ray spectroscopy

Composition of As for each position was measured by Energy-dispersive X-ray spectroscopy (EDS) analysis. As the following High-angle Annular Dark Field Scanning Transmission Electron Microscope (HAADF-STEM) images show, InAsP layers are embedded clearly.

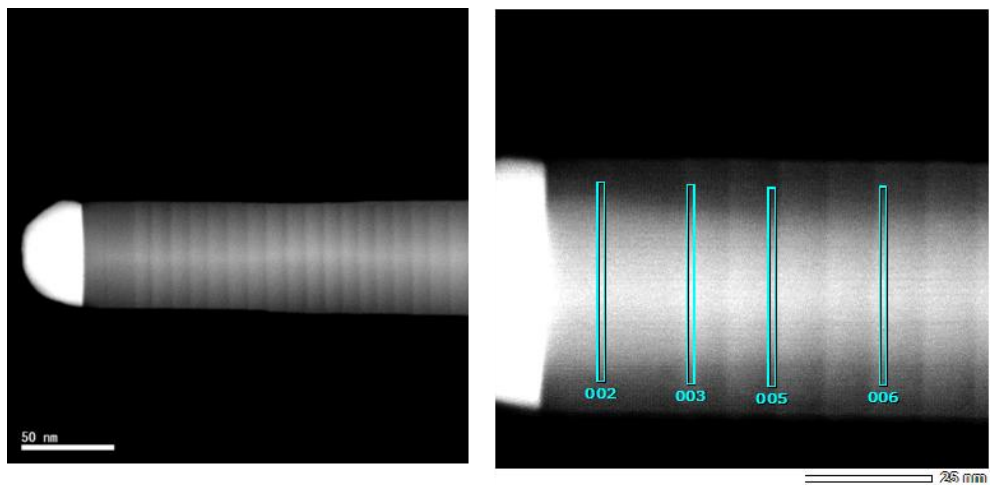

FIG. S1. (a) HAADF-STEM images (scale bar 50 nm). (b) HAADF-STEM images (scale bar 25 nm).

|     | P | As | In | total |  |
| --- | --- | --- | --- | --- | --- |
| 002 | 37.5 | 11.5 | 51.0 | 100.0 | $InAs_{0.23}P_{0.77}$ |
| 003 | 18.7 | 35.7 | 45.6 | 100.0 | $InAs_{0.66}P_{0.34}$ |
| 005 | 24.2 | 31.0 | 44.8 | 100.0 | $InAs_{0.56}P_{0.44}$ |
| 006 | 14.6 | 40.2 | 45.3 | 100.0 | $InAs_{0.73}P_{0.27}$ |

Table. S1. Composition of As for each position

## S2. L-L curve (log scale) and β fitting

Here we replotted the log scale L-L curve from linear curve with spontaneous emission coupling factors (β). The spontaneous emission factor β is obtained by fitting from laser rate equation [1]. Here we estimated Q at transparent condition (κ is determined from Q). We also determined spontaneous emission rate as 0.5 ns from previous pulse measurement [2] and assumed the non-radiative recombination rate is smaller. First we determined threshold from experiment. From this result, we can estimate created carrier. Then, we can estimate carrier density at transparency. After that, we fitted the LL plot using the laser rate equation. Finally we found β is about 0.1. In fact, accuracy of this fitting method is not ideal, however it allows us to estimate approximate value.

## S3. L-L curve (PRBS pump and CW pump)

We measured the L-L curve by using a modulated signal and compared it with the L-L curve obtained by CW pumping and shown in the following figure. The x-axis in the figure is the average power. Because the PRBS pump beam is modulated by an LN modulator without a bias level and 50% of the pump power was attenuated by the LN modulator, the peak power becomes twice the average power. Therefore, the lasing threshold under the CW condition became double that under the PRBS modulated condition.

**REFERNCES**

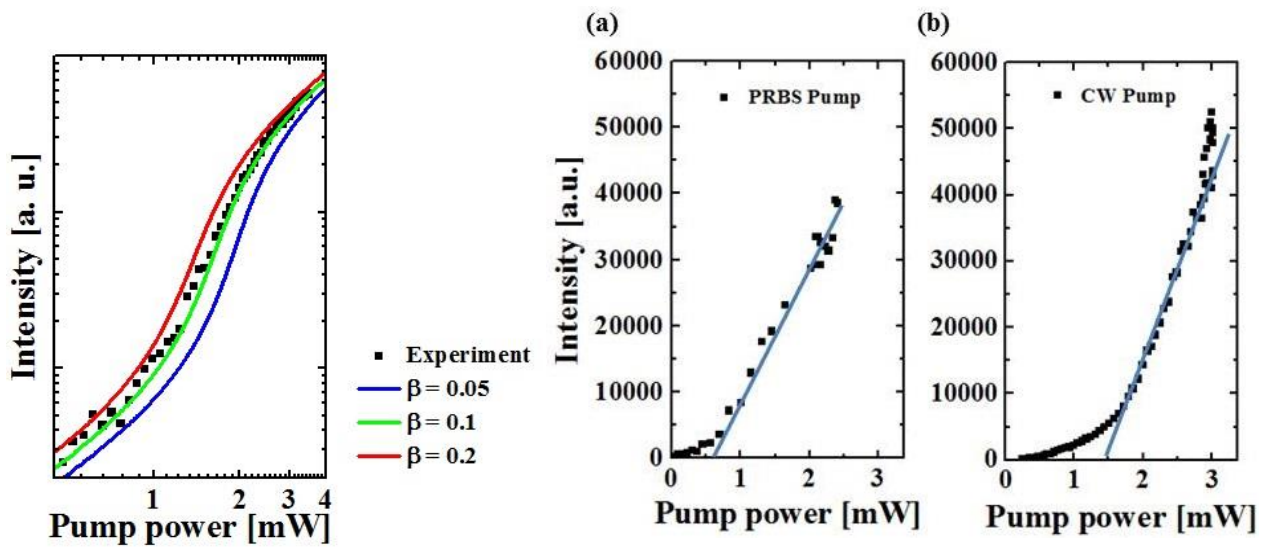

FIG. S2. Log scale L-L curve

FIG. S3. (a) L-L curve obtained by PRBS modulation pumping. (b) L-L curve obtained by CW pumping (no modulation).